\documentclass[11pt,a4paper]{article}

\def\AmSTeX{\leavevmode\hbox{$\mathcal A\kern-.2em\lower.376ex%
        \hbox{$\mathcal M$}\kern-.2em\mathcal S$-\TeX}}
\usepackage{graphicx}
\usepackage{amsmath}        
\usepackage{amssymb}
\usepackage{color}
\usepackage{mathrsfs}       
\usepackage{bm}             
\usepackage[amsmath,thmmarks,hyperref]{ntheorem}
\newif\ifpdf \pdftrue
\ifx\pdfoutput\undefined \pdffalse \fi \ifx\pdfoutput\relax
\pdffalse \fi
\ifx\texonly\undefined\let\texonly\relax\fi
\ifx\endtexonly\undefined\let\endtexonly\relax\fi \texonly
  \let\htmlonly\iffalse
  \let\endhtmlonly\fi
\endtexonly

\htmlonly
  
\endhtmlonly
\texonly
  \usepackage{makeidx}

  \usepackage{multicol}
  

\endtexonly
\title{}
\author{\thanks{}}
\date{}
\begin{document}

\title{Average {speed} and its powers $v^n$ of a heavy quark in quarkonia}

\author{Guo-Li Wang$^{a,b}$, Tai-Fu Feng$^{a,b,c}${\footnote {Corresponding author}}, Xing-Gang Wu$^c$. \\
{\it \small  $^a$ Department of Physics, Hebei University, Baoding
071002, China}\\
{\it \small  $^b$ Hebei Key Laboratory of High-precision Computation and Application} \\ {\it \small of Quantum Field Theory, Baoding
071002, China}\\
{\it \small  $^c$ Department of Physics, Chongqing University, Chongqing 401331, China}}

\maketitle

\baselineskip=20pt
\begin{abstract}
The typical velocity of a heavy quark in a quarkonium is a widely used quantity, in this paper, based on the relativistic Bethe-Salpeter equation method, we calculate
the average values ${\overline{|\boldsymbol{q}|^n}}$ and $ \overline{|\boldsymbol{v}|^n}\equiv v^n$ of a heavy quark in a $S$ wave or $P$ wave quarkonium rest frame, where $\boldsymbol{q}$ and $\boldsymbol{v}$ are the three dimensional momentum and velocity, $n=1,2,3,4$. For a charm quark in $J/\psi$, we obtained $v_{J/\psi}=0.46$, $v^2_{J/\psi}=0.26$, $v^3_{J/\psi}=0.18$, and $v^4_{J/\psi}=0.14$, for a bottom quark in $\Upsilon(1S)$, $v_{\Upsilon(1S)}=0.24$, $v^2_{\Upsilon(1S)}=0.072$, $v^3_{\Upsilon(1S)}=0.025$, and $v^4_{\Upsilon(1S)}=0.010$. The values indicate that ${v^n} >{v^{n_1}}\cdot{v^{n_2}}$, where $n_1+n_2=n$, which is correct for all the charmonia and bottomonia. Our results also show the poor convergence if we make the {speed} expansion in charmonium system, but good for bottomonium. Based on the $v^n$ values and the
following obtained relations $v^n_{4S} > v^n_{3S}> v^n_{2S}>v^n_{1S}$, $v^n_{4P} > v^n_{3P}> v^n_{2P}>v^n_{1P}$ and $v^n_{mP}>v^n_{mS}$ ($n,m=1,2,3,4$), we conclude that highly excited quarkonia have larger relativistic corrections than those of the corresponding low excited and ground states, and there are large relativistic corrections in charmonium system.

\end{abstract}

\section{Introduction}

The heavy quarkonium physics is one of the most hot topics in particle physics after the discovery of $J/\psi$.
Since it is very heavy, heavy quarkonium is a multiscale system which can probe all regimes of quantum chromodynamics (QCD) \cite{brambilla}. So it presents an ideal and unique laboratory for testing the Standard Model and to investigate various aspects of QCD \cite{andronic}.
 It may be crucially important to improve our understanding of QCD
 \cite{brambilla2}.

Because having large masses, the quark and antiquark in heavy quarkonium are expected to move slowly about each other, so the velocity of heavy quark provides a small parameter in which the dynamical scales in heavy quarkonium may be hierarchically ordered and then the corresponding amplitudes where heavy quarkonium is involved in can be systematically expanded in power of velocity of quarks. For example, in the framework of nonrelativistic QCD (NRQCD) which is a powerful effective field theory in describing the quarkonium physics, the $v$ expansion method is widely used, see the paper \cite{bodwin1} for a review. So the typical velocity (as well as the momentum) of heavy quark plays an important role in the physics of quarkonium. Since this typical {speed} could be the expectation or the average value $\overline{v}$, as an expectation value, the relation $\overline{v^n}=\overline{v}^n$ is usually incorrect, so to make {speed} expansion the values of $v^n\equiv\overline{v^n}$ ($n=1,2,3...$) are needed.

The increasing accuracy of the experimental measurements calls for a corresponding accuracy in the theoretical predictions. To increase the theoretical accuracies, relativistic corrections \cite{bodwin5,bodwin6,yjzhang,GK,brambilla1} and the perturbative corrections \cite{petrelli,qiao,qiao1} are usually required.
In the perturbative region, typical $v^n$ values are also needed in some calculations, for example, within the framework of NRQCD, the expansion in $\alpha_s$ always accompanies expansion in $v$, and $\alpha_s(M)\sim v^2$ \cite{bodwin1}.

Considering the relativistic effects, the typical values $v^n$ are widely used. First, without specific calculation, $v^n$ values can be used to give a rough estimation of relativistic effect. If $v$ and all its powers $v^n$ are small, we can conclude that the corresponding relativistic corrections are small, other wise, large relativistic corrections will be obtained. For example, in literature, $v^2_{J/\psi} \approx 0.3$ and $v^2_{\Upsilon}\approx 0.1$ are always cited, so $v^2_{J/\psi}\gg v^2_{\Upsilon}$, then we know that $J/\psi$ has much larger relativistic corrections than $\Upsilon(1S)$ dose. Second, in precise calculation with relativistic corrections, accurate $v^n$ values are widely needed, for example, in the NRQCD method \cite{chung,xiao,bodwin7,jia,chao4}, light-cone method \cite{wangwei}, potential models \cite{ebert,ebert1}, and lattice QCD \cite{bodwin8}, etc.
Third, large relativistic corrections have been found in double-heavy mesons \cite{wangwei,me3,zhu}, especially in highly excited charmonia \cite{me3,me4}, so precise $v^n$ values are more and more important in the physics of quarkonium.

The typical {speeds} of heavy quarks in a ground quarkonium have been studied by different methods, for example, potential models \cite{buchmuller,bodwin4}, calculation using the equation of binding energy or kinetic energy \cite{zhu}, extracting from experimental data \cite{chao}, computing using the Gremm-Kapustin (GK) relation \cite{GK}, etc, but most of the results need to improve accuracy and reduce errors. On the other hand, the knowledge of $v^n$ in excited quarkonium is very limited, however, more and more attentions are paid to the
excited quarkonia \cite{braaten1,bodwin2,nayak,bodwin3,braaten,fazio,chao1}. Because of the shortage of $v^n$ information in excited quarkonium, authors like to choose the same values for excited quarkonium as for the ground state, but this may cause large errors, especially in charmonium and highly excited states, because highly excited states may have larger relativistic corrections than the low excited and ground states, so bigger $v^n$ values should be obtained in highly excited states.

In this paper, using the Bethe-Salpeter (BS) equation method \cite{Salpeter1}, we will calculate the average values of $q^n$ and $v^n$ for a heavy quark in different quarkonia. The motivation is to provide a precise calculation.
It is well know that the BS equation or its reduced version, Salpeter equation \cite{Salpeter2}, is a relativistic equation describing bound state. By solving it, we will obtain relativistic wave function for bound  state, from which we can make precise calculations of $q^n$ and $v^n$ where the relativistic corrections are considered very well, this is most important for highly excited quarkonium because it may has larger relativistic corrections.

The remainder of this paper is organized as follows: Sec. 2 contains a brief review on the BS equation and Salpeter equation. In Sec. 3, we first give the wave functions for various $J^{PC}$ quarkonium bound states, then calculate the average values $q^n$ and $v^n$ (n=1,2,3,4) for a heavy quark inside a quarkonium.
Section 4 is devoted to numerical results and discussions.

\section{The Bethe-Salpeter equation and Salpeter equation}

A quark and an antiquark are bound to a meson by strong interaction, which can be described by the Schrodinger equation if the meson is a nonrelativistic system, but if it is relativistic, then the BS equation \cite{Salpeter1} should be used, because it is a relativistic dynamic equation describing a bound state. For a meson, which containing a quark $1$ and an antiquark $2$, its BS equation can be read as \cite{Salpeter1}
\begin{eqnarray}\label{eq1}
(\not\!p_{1}-m_{1})\chi(P,q)(\not\!p_{2}+m_{2})=i\int\frac{d^{4}k}{(2\pi)^{4}}V(P,k,q)\chi(P,k),
\end{eqnarray}
where $\chi(P,q)$ is the relativistic four dimensional wave function of the meson, $V(P,k,q)$ is the interaction kernel between quark and antiquark. $P$ is the total momentum of the meson, $p_{1}$ and $p_{2}$ are the momenta of the quark and antiquark,
$m_{1}$ and $m_{2}$ ($=m_1$ for quarkonium) are the constituent masses of the quark
and antiquark respectively. $q$ ($k$) is the relative momentum, for a quarkonium which can be defined by the following relations,
$$p_1 = 0.5~ P +
q,\   p_2 = 0.5~ P-q.$$

{ The full BS equation is very complicated, we have to make approach to solve it.
Salpeter equation is the instantaneous version of BS equation, because of including heavy mass, the instantaneous approach is a good method for heavy meson, especially for heavy quarkonium. Refs. \cite{qiao3,ebert3} proved this conclusion by showing a small retardation effect in heavy quarkonium, so in this paper we will solve the Salpeter equation instead of BS equation.}

In the instantaneous approach and in the center of mass system (CMS) of the quarkonium, which is also its rest frame, $P=(M,\boldsymbol{0})$, and the interaction kernel $V(P,k,q)$ becomes to $V(\boldsymbol k, \boldsymbol q)\equiv V(\boldsymbol k - \boldsymbol q)$, then the BS wave function $\chi(P,q)$ becomes to the Salpeter wave function $\varphi(\boldsymbol q)$ after integrating over $q_0$,
\begin{eqnarray}\label{eq2}
 \varphi(\boldsymbol q)\equiv i\int
\frac{\mathrm{d}q_{0}}{2\pi}\chi
(P,q_0,\boldsymbol q).
\end{eqnarray}
With a shorthand symbol
$$\eta(\boldsymbol q)\equiv\int
\frac{\mathrm{d}^3 k}{(2\pi)^{3}}V(\boldsymbol k, \boldsymbol q)\varphi(\boldsymbol q),$$
BS Eq. (\ref{eq1}) can be changed to
\begin{equation}\label{eq3}
\chi(P,q_0,\boldsymbol q)=S_{1}(p_{1})\eta(\boldsymbol q)S_{2}(p_{2}),
\end{equation}
where the propagators can be decomposed into two terms:
\begin{equation}
S_{i}(p_{i})=\frac{\Lambda^{+}_{i}(\boldsymbol q)}{J(i)q_{0}
+0.5M-\omega_{i}+i\epsilon}+
\frac{\Lambda^{-}_{i}(\boldsymbol q)}{J(i)q_{0}+0.5M+\omega_{i}-i\epsilon}\;,
\label{eq4}
\end{equation}
with
$$
\omega_{i}=\sqrt{m_{i}^{2}+{\boldsymbol q}^{2}}\;,\;\;\;
\Lambda^{\pm}_{i}(\boldsymbol q)= \frac{1}{2\omega_{i}}\left[
\gamma_0\omega_{i}\pm
J(i)(m_{i}-{\boldsymbol {q}\cdot \boldsymbol{\gamma}})\right]\;,
$$
where except the imaginary number $i\epsilon$, $i=1, 2$ for quark and antiquark respectively, and
$J(i)=(-1)^{i+1}$. The projection operators $\Lambda^{\pm}_{i}(\boldsymbol q)$ satisfy the following relations:
$$\Lambda^{+}_{i}(\boldsymbol q)+\Lambda^{-}_{i}(\boldsymbol q)=\gamma_0~,\;\;
\Lambda^{\pm}_{i}(\boldsymbol q)\gamma_0
\Lambda^{\pm}_{i}(\boldsymbol q)=\Lambda^{\pm}_{i}(\boldsymbol q)~,\;\;
\Lambda^{\pm}_{i}(\boldsymbol q)\gamma_0
\Lambda^{\mp}_{i}(\boldsymbol q)=0~.$$
After we take the integration over $q_0$ in Eq.
(\ref{eq3}) on both sides, then we get the Salpeter equation,
\begin{equation}\label{eq5}
\varphi(\boldsymbol q)=\frac{\Lambda_{1}^{+}(\boldsymbol q)\eta(\boldsymbol q)
\Lambda_{2}^{+}(\boldsymbol q)}{M-\omega_{1}-\omega_{2}}-\frac{\Lambda_{1}^{-}(\boldsymbol q)\eta(\boldsymbol q)
\Lambda_{2}^{-}(\boldsymbol q)}{M+\omega_{1}+\omega_{2}}.
\end{equation}

If we introduce the notations
$$\varphi^{\pm\pm}(\boldsymbol q)\equiv\Lambda_{1}^{\pm}(\boldsymbol q)\gamma_0\varphi(\boldsymbol q)\gamma_0\Lambda_{2}^{\pm}(\boldsymbol q),$$
the Salpeter wave function can be separated into four terms,
\begin{equation}\label{eq6}
\varphi^{\pm\pm}(\boldsymbol q)\equiv \varphi^{++}(\boldsymbol q)+\varphi^{+-}(\boldsymbol q)+\varphi^{-+}(\boldsymbol q)+\varphi^{--}(\boldsymbol q),
\end{equation}
where $\varphi^{++}(\boldsymbol q)$ is the positive wave function, $\varphi^{--}(\boldsymbol q)$ is the negative one.

Using the relations of projection operators, the Salpeter equation can be written as \cite{Salpeter2}
\begin{eqnarray}\label{eq7}
&&(M-\omega_{1}-\omega_{2})\varphi(\boldsymbol q)^{++}=\Lambda_{1}^{+}(\boldsymbol q)\eta(\boldsymbol q)
\Lambda_{2}^{+}(\boldsymbol q),\nonumber \\
&&(M+\omega_{1}+\omega_{2})\varphi(\boldsymbol q)^{--}=-\Lambda_{1}^{-}(\boldsymbol q)\eta(\boldsymbol q)
\Lambda_{2}^{-}(\boldsymbol q),\nonumber \\
&&\varphi(\boldsymbol q)^{+-}=0,~~ \varphi(\boldsymbol q)^{-+}=0.
\end{eqnarray}
Since $\omega_{1}$ and $\omega_{2}$ are the energies of quark and antiquark inside a quarkonium, the value of $\omega_{1}+\omega_{2}$ is close to the quarkonium mass $M$, then in Salpeter equation, the quantity of $(M-\omega_{1}-\omega_{2})$ is much smaller than $(M+\omega_{1}+\omega_{2})$, so one can conclude that the value of $\varphi(\boldsymbol q)^{++}$ is much larger than that of $\varphi(\boldsymbol q)^{--}$, so in literature, usually only the first equation is solved instead of the whole four equations. But we point out that this will lose the benefit of Sapeter equation, so we should solve the full Salpeter equation to obtain a relativistic wave function of a quarkonium.

In our method, the Cornell potential which is a linear scalar
potential plus a Coulomb vector potential, is chosen as the
instantaneous interaction kernel $V$,
\begin{eqnarray}\label{eq8}
V(r)=\lambda r + V_0-\frac{4}{3}\frac{\alpha_s}{r}.
\end{eqnarray}

\section{Relativistic calculation of the average values ${q}^n$ and ${v}^n$ of a heavy quark in a quarkonium}

We adopt the classification of $Q\bar Q$ quarkonium in terms of the radial
quantum number $n$, the spin $S$, the orbital angular momentum $L$ and the total angular momentum $J$. Then state identified by $n^{2S+1}L_J$ corresponds to a meson, in this paper, we consider two $S$ wave states, pseudoscalar $^1S_0$ and vector $^3S_1$, four $P$ wave states, $^1P_1$, $^3P_0$, $^3P_1$, and $^3P_2$. Equally, we can also use the $J^{PC}$ to identify the states, where $P=(-1)^{L+1}$ is the parity and $C=(-1)^{L+S}$ the charge-conjugation parity. So two $S$ wave quarkonia can be labeled as $0^{-+}$ and $1^{--}$, four $P$ wave quarkonia can be labeled as $1^{+-}$, $0^{++}$, $1^{++}$ and $2^{++}$, correspondingly.

The relativistic wave functions with certain quantum numbers
$0^{-+}(^1S_0)$, $1^{--}(^3S_1)$, $1^{+-}(^1P_1)$, $0^{++}(^3P_0)$, $1^{++}(^3P_1)$ and $2^{++}(^3P_2)$ can be written as \cite{me1,me5,me6,me7}
\begin{eqnarray}\label{eq9}
\varphi_{0^{-+}}(\boldsymbol q)&=&M\left[\gamma_0 a_1(\boldsymbol q)+{a}_{2}(\boldsymbol q)+\frac{\boldsymbol {q}\cdot\boldsymbol {\gamma}~ \gamma_0}{M}{a}_{3}(\boldsymbol q)\right]{\gamma}_5,
\notag\\
{\varphi}_{1^{--}}({\boldsymbol q})&=&(-\boldsymbol {q}\cdot \boldsymbol {\epsilon})\left[b_1({\boldsymbol q})
-\frac{\boldsymbol {q}\cdot\boldsymbol {\gamma}}{M}b_3({\boldsymbol q})+\frac{\boldsymbol {q}\cdot\boldsymbol {\gamma}~ \gamma_0}{M}b_4({\boldsymbol q})\right]-M{\boldsymbol \epsilon}\cdot{\boldsymbol \gamma}~b_5({\boldsymbol q})\notag\\
&&+M \gamma_0{\boldsymbol \epsilon}\cdot{\boldsymbol \gamma}~b_6({\boldsymbol q})
+\gamma_0({\boldsymbol \epsilon}\cdot{\boldsymbol \gamma} ~{\boldsymbol {q}\cdot\boldsymbol {\gamma}}+\boldsymbol {q}\cdot \boldsymbol {\epsilon})b_8({\boldsymbol q}),\notag\\
{\varphi}_{0^{++}}({\boldsymbol q})&=&-\boldsymbol {q}\cdot\boldsymbol {\gamma}~{f}_{1}({\boldsymbol q})+{\boldsymbol {q}\cdot\boldsymbol {\gamma}~ \gamma_0}~{f}_{2}({\boldsymbol q})
+M{f}_{3}({\boldsymbol q}),\notag\\
{\varphi}_{1^{++}}({\boldsymbol q})&=&i{\varepsilon}_{0\mu\alpha\beta}q^{\alpha}{\epsilon}^{\beta}
\biggl[{\gamma}^{\mu}g_1({\boldsymbol q})+\gamma^0{\gamma}^{\mu}g_2({\boldsymbol q})+ig_4({\boldsymbol q}){\varepsilon}^{0\mu\rho\delta}q_{\rho}{\gamma}_{\delta}{\gamma}_5/M\biggl],\notag\\
{\varphi}_{1^{+-}}({\boldsymbol q})&=&\boldsymbol {q}\cdot \boldsymbol {\epsilon}\left[h_1({\boldsymbol q})+\gamma_0h_2({\boldsymbol q})+\frac{\boldsymbol {q}\cdot\boldsymbol {\gamma}~ \gamma_0}{M}h_4({\boldsymbol q})\right]{\gamma}_5,\notag\\
\varphi_{2^{++}}(\boldsymbol q)&=&
{\varepsilon}_{\mu\nu}{q^{\nu}}
\biggl\{{q^{\mu}}\left[j_1(\boldsymbol q)-
\frac{\boldsymbol {q}\cdot\boldsymbol {\gamma}}{M}j_3(\boldsymbol q)+\frac{\boldsymbol {q}\cdot\boldsymbol {\gamma}~ \gamma_0}{M}j_4({\boldsymbol q})\right] \notag\\ &&+
M{\gamma^{\mu}}\left[j_5(\boldsymbol q)+ {\gamma^0}j_6(\boldsymbol q)\right]-{i}
\epsilon^{0\mu\beta\gamma}
q_{\beta}\gamma_{\gamma}\gamma_{5}j_8(\boldsymbol q)\biggl\},
\end{eqnarray}
where the radial wave functions $a_i(\boldsymbol q),b_i(\boldsymbol q),f_i(\boldsymbol q),g_i(\boldsymbol q),h_i(\boldsymbol q)$ and
$j_i(\boldsymbol q)$ are functions of $\boldsymbol q^2$, so there is no $\boldsymbol q^2$ terms in Eq. (\ref{eq9}). There is also no $P\cdot q$ terms because in the instantaneous approximation $q=(0,\boldsymbol q)$ and $P\cdot q=0$. $\boldsymbol \epsilon$ is the polarization vector of a $1^{--}$, $1^{++}$ or $1^{+-}$ state, $\epsilon_{\mu\nu}$ is the polarization tensor of the $2^{++}$ state. With these wave
function forms, we solved the Salpeter Eq.~(\ref{eq7}) and obtained the mass spectra and numerical values of wave functions. The details of how to solve the full Salpeter equations can be found in our previous papers \cite{me2}.

The normalization conditions for above wave functions are \cite{me2},
\begin{eqnarray}\label{eq10}
&&\int\frac{\mathrm{d}^3{q}}{(2\pi)^3}2a_1
a_2M\left[
\frac{\omega_{1}}{m_{1}}+\frac{m_{1}}{\omega_{1}}
+\frac{{\boldsymbol q}^2}{\omega_{1}m_{1}} \right]=1~,\nonumber \\
&&\int \frac{\mathrm{d}^3{q}}{(2\pi)^3}\frac{4\omega_1}{3m_1M}\left[
3b_5b_6{M^2}+
b_4b_5{\boldsymbol q}^2-b_3{\boldsymbol q}^2\left(b_4\frac{ {\boldsymbol q}^2}{M^2}+b_6\right) \right]=1~,\nonumber \\
&&\int \frac{\mathrm{d}^3{q}}{(2\pi)^3}\frac{4f_1f_2
\omega_1{{\boldsymbol q}^2}}{m_1M}=1~,\nonumber \\
&&\int \frac{\mathrm{d}^3{q}}{(2\pi)^3}\frac{8g_1g_2
\omega_1{\boldsymbol q}^2}{3m_1M}=1~,\nonumber \\
&&\int \frac{\mathrm{d}^3{q}}{(2\pi)^3}\frac{4h_1h_2
\omega_1{\boldsymbol q}^2}{3m_1M}=1~,\nonumber \\
&&\int\frac{\mathrm{d}^3{q}}{(2\pi)^3}\frac{4 \omega_1 {\boldsymbol q}^2}{15m_1M}\left[
5j_5j_6M^2 +2j_4j_5 {{\boldsymbol q}}^2 -2
{\boldsymbol q}^2j_3\left(j_4\frac{{\boldsymbol q}^2}{M^2}+j_6\right) \right]=1~.
\end{eqnarray}

{ In the CMS of the quarkonium, we have the relation ${\boldsymbol p_1}={\boldsymbol q}={-\boldsymbol p_2}$, so $\boldsymbol q$ is the quark momentum.
The normalization conditions can be summarized as $\int dq f^2(q)=1$, which means the probability we find the quark in the whole momentum space is unity, and $f^2(q)dq$ is the possibility that the quark momentum takes on the values $q\to q+dq$, so same to the method of Maxwell speed distribution, we define the average value, $\langle {q}^n \rangle \equiv\overline {{|\boldsymbol q| }^n}$, for a quark inside quarkonium, which can be calculated as followings,}
\begin{equation}\label{eq11}
\langle { q}^n \rangle_{0^{-+}} =\int\frac{\mathrm{d}^3{q}}{(2\pi)^3}2a_1
a_2{|\boldsymbol q|}^nM\left\{
\frac{\omega_{1}}{m_{1}}+\frac{m_{1}}{\omega_{1}}
+\frac{{\boldsymbol q}^2}{\omega_{1}m_{1}} \right\},
\end{equation}
\begin{equation}\label{eq12}
\langle { q}^n \rangle_{1^{--}}=\int \frac{\mathrm{d}^3{q}}{(2\pi)^3}\frac{4\omega_1 {|\boldsymbol q|}^n}{3m_1M}\left[
3b_5b_6{M^2}+
b_4b_5{\boldsymbol q}^2-b_3{\boldsymbol q}^2\left(b_4\frac{ {\boldsymbol q}^2}{M^2}+b_6\right) \right],
\end{equation}
\begin{equation}\label{eq13}
\langle { q}^n \rangle_{0^{++}}=\int \frac{\mathrm{d}^3{q}}{(2\pi)^3}\frac{4f_1f_2
\omega_1{{|\boldsymbol q|}^{2+n}}}{m_1M},
\end{equation}
\begin{equation}\label{eq14}
\langle { q}^n \rangle_{1^{++}}=\int \frac{\mathrm{d}^3{q}}{(2\pi)^3}\frac{8g_1g_2
\omega_1{|\boldsymbol q| }^{2+n}}{3m_1M},
\end{equation}
\begin{equation}\label{eq15}
\langle { q}^n \rangle_{1^{+-}}=\int \frac{\mathrm{d}^3{q}}{(2\pi)^3}\frac{4h_1h_2
\omega_1{|\boldsymbol q| }^{2+n}}{3m_1M},
\end{equation}
\begin{equation}\label{eq16}
\langle { q}^n \rangle_{2^{++}}=\int\frac{\mathrm{d}^3{q}}{(2\pi)^3}\frac{4 \omega_1 {|\boldsymbol q| }^{2+n}}{15m_1M}\left\{
5j_5j_6M^2 +2j_4j_5 {{\boldsymbol q}}^2 -2
{\boldsymbol q}^2j_3\left(j_4\frac{{\boldsymbol q}^2}{M^2}+j_6\right) \right\},
\end{equation}
{ where $|\boldsymbol q|$ is the absolute magnitude of momentum.
The calculated method of average value shows us obviously that it is also the expectation value, so we have the relation}
\begin{equation}\label{eq17}
{q}^n \equiv \overline {{|\boldsymbol q| }^n}\equiv \langle { q}^n \rangle ,~~v^n \equiv \overline {{ |\boldsymbol v|}^n}\equiv \langle{ v}^n\rangle  ,
\end{equation}
where $\boldsymbol v=\frac{\boldsymbol q}{m_1}$ is the quark velocity.

\section{Numerical results and discussions}

When solving the full Salpeter equations, we choose the same parameter values as in the paper \cite{me2}, which are determined by fitting mass spectra of charmonia and bottomonia, and the quark masses are chosen as
$m_c=1.62$ GeV and $m_b=4.96$ GeV. Using the Eqs. (\ref{eq11}-\ref{eq16}), the expectation values of $q^n$ and $v^n$ for a heavy quark inside different quarkonia are calculated, and results are shown in Tables \ref{tab1}~-~\ref{tab4}, in these tables, we also show the mass spectra where the masses of ground states are input.

{ In Cornell potential, at large momentum, the interaction between quarks is dominated by the Coulomb potential. When calculating $q^n$ or $v^n$,
with the increase of $n$, Bodwin et al. [26], found the problem of ultraviolet divergence, we meet the same problem when $n\geq 5$, but we did not make use of hard-cutoff regulator to do the calculation like they did, only show the stable results of $q^n$ and $v^n$ where $n\leq 4$.

In the numerical calculation, limited by computing power, we have to make hard cutoff of integration variable, relative momentum $q$, we find if we choose $q_{max}=3.87~GeV$, the numerical results are stable. The physical reason we can make hard cutoff is that just as this paper shows, the probability of heavy quark inside quarkonium with large momentum or speed is very small, so the value of radial wave function tends to zero at large momentum. To investigate the ultraviolet behavior of wave functions and the stability of the results, we vary the cutoff $q_{max}$ and give the relative increasing of $v^n$ where $n=,3,4,5,6$. When we choose $q_{max}=7.71~GeV$, the increasing of $\{v^3,v^4,v^5,v^6\}$ for $\eta_c(1S)$ are $\{2.9\%,8.9\%,25\%,49\%\}$, for $\eta_c(2S)$ $\{2.1\%,5.7\%,14\%,31\%\}$, for $\eta_c(3S)$ $\{1.9\%,4.4\%,11\%,25\%\}$. We can see that, the convergence is not good for $v^5$, and bad for $v^6$, so in this paper, we only show the results of $q^n$ and $v^n$ where $n\leq 4$ using $q_{max}=3.87~GeV$.}

\begin{table} \caption{Average values of ${q}^n$ and ${v}^n$ of charm quark inside $0^{-+}$ and $1^{--}$ charmonia, where the masses of $m_{\eta_c(1S)}=2980.3$ MeV and $m_{J/\psi}=3096.9$ MeV are input.}
\begin{center}
\begin{tabular}{|c|c|c|c|c|c|c|c|c|c|} \hline\hline
State&Mass&~${q}$~&~${q}^2$~&~${q}^3$~&~${q}^4$~&
${v}$&${v}^2$&${v}^3$&${v}^4$\\\hline\hline
~$\eta_c(1S)$~&~2980.3~&~0.728~&~0.653~&~0.706~&~0.915~&~0.449~&~0.249~&~0.166~&~0.133~\\\hline
~$\eta_c(2S)$~&~3576.4~&~0.796~&~0.885~&~1.17~&~1.72~&~0.491~&~0.337~&~0.274~&~0.249~\\\hline
~$\eta_c(3S)$~&~3948.8~&~0.877~&~1.06~&~1.52~&~2.41~&~0.541~&~0.405~&~0.357~&~0.350~\\\hline
~$\eta_c(4S)$~&~4224.6~&~0.939~&~1.21~&~1.82~&~3.03~&~0.579~&~0.460~&~0.428~&~0.440~\\\hline\hline

~$J/\psi$~&~3096.9~&~0.743~&~0.679~&~0.744~&~0.970~&~0.459~&~0.259~&~0.175~&~0.141~\\\hline
~$\psi(2S)$~&~3688.1~&~0.810~&~0.914~&~1.22~&~1.81~&~0.500~&~0.348~&~0.286~&~0.262~\\\hline
~$\psi(3S)$~&~4056.8~&~0.894~&~1.10~&~1.59~&~2.54~&~0.552~&~0.419~&~0.374~&~0.369~\\\hline
~$\psi(4S)$~&~4329.4~&~0.956~&~1.25~&~1.90~&~3.19~&~0.590~&~0.476~&~0.447~&~0.463~\\\hline\hline
\end{tabular}\label{tab1}
\end{center}
\end{table}

Table \ref{tab1} shows the average values $q^n$ and $v^n$ of a charm quark inside pseudoscalars $\eta_c(1S-4S)$ and vectors $\psi(1S-4S)$. In cases of $\{\eta_c(1S),J/\psi\}$, $v=\{0.45,0.46\}$, $v^2=\{0.25,0.26\}$, $v^3=\{0.17,0.18\}$, $v^4=\{0.13,0.14\}$, so approximately we have $v^n_{\eta_c(mS)}\approx v^n_{\psi(mS)}$ ($n,m=1,2,3,4$), this could be a double check of the correctness of this model since in a nonrelativistic model, they are treated as same values, and the difference between them comes from the corrections of order $v^2$ \cite{bodwin1}. Our results in Table \ref{tab1} indicate that the average value $v^n$ in a highly excited state is larger than in a low excited state, that is we have the relation $v^n_{4S}>v^n_{3S}>v^n_{2S}>v^n_{1S}$ ($n=1,2,3,4$), for example, $v^2_{\psi(4S)}=0.48>v^2_{\psi(3S)}=0.42>v^2_{\psi(2S)}=0.35>v^2_{J/\psi}=0.26$.

Reference \cite{buchmuller} using potential model predicted the velocity squared $v^2$ of $\psi$ system, their results are $v^2_{J/\psi}=0.23$, $v^2_{\psi(2S)}=0.29$, $v^2_{\psi(3S)}=0.36$ and $v^2_{\psi(4S)}=0.44$, which are comparable with ours. Also based on potential model, Ref. \cite{bodwin4} predicted $\langle v^2_{J/\psi}\rangle=0.25\pm0.05\pm0.08$, where $\langle v^2_{J/\psi} \rangle$ is not the expectation value defined in this paper, but the long distance matrix element of $J/\psi$ \cite{bodwin,chao}, while based on NRQCD velocity-scaling rules \cite{bodwin1} it is equal approximately to $\overline {v^2}$ \cite{bodwin}. We can see that, their value
$\langle v^2_{J/\psi} \rangle$ is very consistent with ours. Also based on potential model that employs Cornell potential, Ref. \cite{bodwin} obtained $\langle v^2_{J/\psi}\rangle=0.224$ and $\langle v^2_{\eta_c(1S)}\rangle=0.226$.
In Ref. \cite{chao}, with the experimental $\gamma\gamma$ width $\Gamma^{\gamma\gamma}_{{\eta}_{c}}$ as input, they obtained $\langle v^2_{\eta_c(1S)} \rangle=0.228^{+0.126}_{-0.100}$, if using the total width $\Gamma^{total}_{\eta_c(1S)}$ as input, their result is $\langle v^2_{\eta_c(1S)}\rangle=0.234^{+0.121}_{-0.099}$, all above predictions are close to ours.

\begin{table} \caption{Average values of ${q}^n$ and ${v}^n$ of charm quark inside $0^{++}$, $1^{++}$, $2^{++}$ and $1^{+-}$ charmonia, where the masses of ground $1P$ states are input.}
\begin{center}
\begin{tabular}{|c|c|c|c|c|c|c|c|c|c|} \hline\hline
State&Mass&~${q}$~&~${q}^2$~&~${q}^3$~&~${q}^4$~&
${v}$&${v}^2$&${v}^3$&${v}^4$\\\hline
~$\chi_{c0}(1P)$~&~3414.7~&~0.838~&~0.796~&~0.850~&~1.02~&~0.517~&~0.303~&~0.200~&~0.148~\\\hline
~$\chi_{c0}(2P)$~&~3836.8~&~0.882~&~0.992~&~1.30~&~1.87~&~0.544~&~0.378~&~0.305~&~0.271~\\\hline
~$\chi_{c0}(3P)$~&~4140.1~&~0.937~&~1.15~&~1.64~&~2.58~&~0.579~&~0.439~&~0.387~&~0.375~\\\hline
~$\chi_{c0}(4P)$~&~4376.9~&~0.985~&~1.28~&~1.94~&~3.21~&~0.608~&~0.489~&~0.455~&~0.466~\\\hline\hline

~$\chi_{c1}(1P)$~&~3510.3~&~0.849~&~0.814~&~0.874~&~1.05~&~0.524~&~0.310~&~0.205~&~0.152~\\\hline
~$\chi_{c1}(2P)$~&~3928.7~&~0.896~&~1.02~&~1.34~&~1.93~&~0.553~&~0.388~&~0.315~&~0.280~\\\hline
~$\chi_{c1}(3P)$~&~4228.8~&~0.953~&~1.18~&~1.70~&~2.68~&~0.588~&~0.451~&~0.401~&~0.389~\\\hline
~$\chi_{c1}(4P)$~&~4463.1~&~1.00~&~1.32~&~2.01~&~3.34~&~0.619~&~0.503~&~0.473~&~0.485~\\\hline
\hline
~$\chi_{c2}(1P)$~&~3555.6~&~0.839~&~0.791~&~0.829~&~0.959~&~0.518~&~0.301~&~0.195~&~0.139~\\\hline
~$\chi_{c2}(2P)$~&~3971.0~&~0.896~&~1.01~&~1.30~&~1.84~&~0.553~&~0.385~&~0.307~&~0.267~\\\hline
~$\chi_{c2}(3P)$~&~4269.3~&~0.957~&~1.18~&~1.68~&~2.60~&~0.590~&~0.451~&~0.396~&~0.378~\\\hline
~$\chi_{c2}(4P)$~&~4502.0~&~1.01~&~1.33~&~2.00~&~3.28~&~0.622~&~0.505~&~0.471~&~0.476~\\\hline
\hline
~$h_c(1P)$~&~3526.0~&~0.844~&~0.802~&~0.851~&~1.00~&~0.521~&~0.306~&~0.200~&~0.146~\\\hline
~$h_c(2P)$~&~3943.0~&~0.896~&~1.01~&~1.32~&~1.89~&~0.553~&~0.387~&~0.311~&~0.274~\\\hline
~$h_c(3P)$~&~4242.4~&~0.955~&~1.18~&~1.69~&~2.64~&~0.589~&~0.451~&~0.398~&~0.384~\\\hline
~$h_c(4P)$~&~4476.2~&~1.00~&~1.32~&~2.01~&~3.31~&~0.620~&~0.504~&~0.472~&~0.481~\\\hline
\end{tabular}\label{tab2}
\end{center}
\end{table}

The average values $q^n$ and $v^n$ of a charm quark inside $P$ wave charmonia are shown in Table \ref{tab2}. First, we have the relations $v^n_{\chi_{c0}(mP)}\approx v^n_{\chi_{c1}(mP)}\approx v^n_{\chi_{c2}(mP)}\approx v^n_{h_{c}(mP)}$ ($n,m=1,2,3,4$), for example, $v^2_{\chi_{c0}(1P)}\approx v^2_{\chi_{c1}(1P)}\approx v^2_{\chi_{c2}(1P)}\approx v^2_{h_{c}(1P)}=0.30$, this also can be as a double check that the method is correct because in a nonrelativistic limit one use a same wave function for these four states, but we use four different wave functions and normalization conditions for them, while we obtained similar results. Second, similar to $S$ wave results, there are the relations $v^n_{\chi_{cJ}(4P)}>v^n_{\chi_{cJ}(3P)}>v^n_{\chi_{cJ}(2P)}>v^n_{\chi_{cJ}(1P)}$ and $v^n_{h_{c}(4P)}>v^n_{h_{c}(3P)}>v^n_{h_{c}(2P)}>v^n_{h_{c}(1P)}$ ($n=1,2,3,4$, $J=1,2,3$). Third, compared with the corresponding $S$ wave state, we have the relations $v^n_{\chi_{cJ}(mP)}>v^n_{\psi(mS)}$ ($n,m=1,2,3,4$, $J=1,2,3$) and $v^n_{h_{c}(mP)}>v^n_{\eta_{c}(mS)}$ ($n,m=1,2,3,4$), for example, $v^2_{\chi_{c1}(1P)}=0.31>v^2_{\psi(1S)}=0.26$, $v^2_{h_{c}(2P)}=0.39>v^2_{\eta_{c}(2S)}=0.34$, so the usually used relations $v^n_{\chi_{cJ}(mP)}=v^n_{\psi(mS)}$ and $v^n_{h_{c}(mP)}=v^n_{\eta_{c}(mS)}$ are incorrect.

Reference \cite{buchmuller} predicted $v^2_{c\bar c(1P)}=0.25$ and $v^2_{c\bar c(2P)}=0.32$, which are comparable with ours, but a little smaller. Within QCD sum rules, Ref. \cite{braguta} predicted $v^2_{c\bar c(1P)}=0.30\pm0.10$ and $v^4_{c\bar c(1P)}=0.12\pm0.04$, Ref. \cite{hwang} using the light-front framework given $v^2_{c\bar c(1P)}=0.317$ and $v^4_{c\bar c(1P)}=0.118$, these two results are very close to ours.

We find that in charmonium system, see Table \ref{tab1} and Table \ref{tab2}, the results indicate the poor convergence if we make the velocity expansion (Ref. \cite{chen} got a similar conclusion), especially for highly excited states, where the convergence is vary bad. For example, we get $v_{J/\psi}=0.46$, $v^2_{J/\psi}=0.26$, $v^3_{J/\psi}=0.18$ and $v^4_{J/\psi}=0.14$,  the convergence rate on the power of $v$ is very slow, and the values of high power of $v$ are large, both of them indicate there are large relativistic corrections in $J/\psi$. Inside $\psi(4S)$, the values are $v_{\psi(4S)}=0.59$, $v^2_{\psi(4S)}=0.48$, $v^3_{\psi(4S)}=0.45$ and $v^4_{\psi(4S)}=0.46$, the $v$ expansion is very bad in this case, the reason is that there are three nodes in the wave function of $\psi(4S)$. The structure of nodes results in big contribution from large $v$ region, so we obtained big average values $\overline {v^n}$, which indicate very large relativistic corrections in $\psi(4S)$. By comparing the $v^n$ values, we conclude that highly excited states (including radially and orbitally excited state) have larger relativistic corrections than those of low excited and ground states, which make the convergence of the velocity expansion very bad in highly excited states. The authors in Ref. \cite{chao3} also found the velocity expansion in the present NRQCD framework suffers from large high order relativistic corrections in another way which due to ignoring the momentum of soft hadrons.

\begin{table} \caption{Average values of ${q}^n$ and ${v}^n$ of bottom quark inside $0^{-+}$ and $1^{--}$ bottomonia, where the masses of ground $1S$ states are input.}
\begin{center}
\begin{tabular}{|c|c|c|c|c|c|c|c|c|c|} \hline\hline
State&Mass&~${q}$~&~${q}^2$~&~${q}^3$~&~${q}^4$~&
${v}$&${v}^2$&${v}^3$&${v}^4$\\\hline
~$\eta_b(1S)$~&~9390.2~&~1.19~&~1.74~&~3.02~&~6.00~&~0.240~&~0.0708~&~0.0247~&~0.00991~\\\hline
~$\eta_b(2S)$~&~9950.0~&~1.18~&~2.00~&~4.04~&~9.00~&~0.237~&~0.0811~&~0.0331~&~0.0149~\\\hline
~$\eta_b(3S)$~&~10311.4~&~1.21~&~2.11~&~4.43~&~10.3~&~0.244~&~0.0860~&~0.0363~&~0.0170~\\\hline
~$\eta_b(4S)$~&~10554.0~&~1.30~&~2.41~&~5.32~&~13.0~&~0.262~&~0.0981~&~0.0436~&~0.0214~\\\hline
\hline
~$\Upsilon(1S)$~&~9460.5~&~1.20~&~1.76~&~3.05~&~6.10~&~0.241~&~0.0715~&~0.0250~&~0.0101~\\\hline
~$\Upsilon(2S)$~&~10023.1~&~1.16~&~1.96~&~3.95~&~8.79~&~0.234~&~0.0797~&~0.0324~&~0.0145~\\\hline
~$\Upsilon(3S)$~&~10368.9~&~1.25~&~2.22~&~4.74~&~11.1~&~0.251~&~0.0904~&~0.0388~&~0.0184~\\\hline
~$\Upsilon(4S)$~&~10635.8~&~1.34~&~2.52~&~5.61~&~13.7~&~0.270~&~0.103~&~0.0460~&~0.0226~\\\hline
\end{tabular}\label{tab3}
\end{center}
\end{table}

The corresponding results of the average $q^n$ and $v^n$ values of a bottom quark inside a $S$ wave bottomonium are shown in Table \ref{tab3}. Similar to the charm quark case, we have $q^n_{\eta_b(mS)}\approx q^n_{\Upsilon(mS)}$ and $v^n_{\eta_b(mS)}\approx v^n_{\Upsilon(mS)}$ ($n,m=1,2,3,4$). And the average value $v^n_{\Upsilon(mS)}$ is much smaller than the corresponding $v^n_{\psi(mS)}$ in Table \ref{tab1}, for example, $v^2_{\Upsilon(1S)}=0.072\ll v^2_{J/\psi}=0.26$, which indicate there are much smaller relativistic corrections in bottomonium than those in charmonium. We also have the relation $v^n_{4S}>v^n_{3S}>v^n_{2S}>v^n_{1S}$ ($n=1,2,3,4$), except $v_{\eta_b(2S)}=v_{\eta_b(1S)}=0.24$ and $v_{\Upsilon(2S)}=0.23<v_{\Upsilon(1S)}=0.24$, but these deviations cannot change our conclusion, that the relativistic corrections in a highly excited state is larger than that in a low excited state. In Ref. \cite{buchmuller}, the authors predicted $v^2_{\Upsilon(1S)}=0.077$, $v^2_{\Upsilon(2S)}=0.075$, $v^2_{\Upsilon(3S)}=0.085$ and $v^2_{\Upsilon(4S)}=0.098$, which are consistent well with our results $v^2_{\Upsilon(1S)}=0.072$, $v^2_{\Upsilon(2S)}=0.080$, $v^2_{\Upsilon(3S)}=0.090$ and $v^2_{\Upsilon(4S)}=0.10$.

\begin{table} \caption{Average values of ${q}^n$ and ${v}^n$ of bottom quark inside $0^{++}$, $1^{++}$, $2^{++}$ and $1^{+-}$ bottomonia, where the masses of ground $1P$ states are input.}
\begin{center}
\begin{tabular}{|c|c|c|c|c|c|c|c|c|c|} \hline\hline
State&Mass&~${q}$~&~${q}^2$~&~${q}^3$~&~${q}^4$~&
${v}$&${v}^2$&${v}^3$&${v}^4$\\\hline
~$\chi_{b0}(1P)$~&~9859.0~&~1.29~&~1.88~&~3.08~&~5.62~&~0.259~&~0.0764~&~0.0253~&~0.00929~\\\hline
~$\chi_{b0}(2P)$~&~10240.6~&~1.29~&~2.15~&~4.22~&~9.07~&~0.259~&~0.0875~&~0.0346~&~0.0150~\\\hline
~$\chi_{b0}(3P)$~&~10524.7~&~1.35~&~2.44~&~5.20~&~12.1~&~0.271~&~0.0993~&~0.0426~&~0.0200~\\\hline
~$\chi_{b0}(4P)$~&~10757.0~&~1.39~&~2.65~&~5.92~&~14.5~&~0.280~&~0.108~&~0.0485~&~0.0239~\\\hline
\hline
~$\chi_{b1}(1P)$~&~9892.2~&~1.28~&~1.88~&~3.08~&~5.60~&~0.259~&~0.0763~&~0.0252~&~0.00925~\\\hline
~$\chi_{b1}(2P)$~&~10272.7~&~1.29~&~2.16~&~4.25~&~9.13~&~0.260~&~0.0880~&~0.0348~&~0.0151~\\\hline
~$\chi_{b1}(3P)$~&~10556.2~&~1.35~&~2.45~&~5.22~&~12.2~&~0.271~&~0.0994~&~0.0427~&~0.0201~\\\hline
~$\chi_{b1}(4P)$~&~10787.8~&~1.38~&~2.61~&~5.83~&~14.2~&~0.278~&~0.106~&~0.0477~&~0.0235~\\\hline
\hline
~$\chi_{b2}(1P)$~&~9913.3~&~1.26~&~1.80~&~2.88~&~5.13~&~0.254~&~0.0731~&~0.0236~&~0.00847~\\\hline
~$\chi_{b2}(2P)$~&~10284.0~&~1.24~&~2.04~&~3.94~&~8.35~&~0.251~&~0.0830~&~0.0323~&~0.0138~\\\hline
~$\chi_{b2}(3P)$~&~10591.6~&~1.36~&~2.49~&~5.32~&~12.4~&~0.275~&~0.101~&~0.0436~&~0.0205~\\\hline
~$\chi_{b2}(4P)$~&~10786.9~&~1.43~&~2.76~&~6.17~&~15.0~&~0.289~&~0.112~&~0.0506~&~0.0248~\\\hline
\hline
~$h_b(1P)$~&~9900.2~&~1.27~&~1.84~&~2.97~&~5.32~&~0.257~&~0.0747~&~0.0243~&~0.00879~\\\hline
~$h_b(2P)$~&~10280.4~&~1.25~&~2.05~&~3.93~&~8.30~&~0.252~&~0.0832~&~0.0322~&~0.0137~\\\hline
~$h_b(3P)$~&~10562.0~&~1.34~&~2.42~&~5.11~&~11.8~&~0.270~&~0.0983~&~0.0419~&~0.0195~\\\hline
~$h_b(4P)$~&~10793.8~&~1.39~&~2.65~&~5.90~&~14.4~&~0.281~&~0.108~&~0.0484~&~0.0237~\\\hline
\end{tabular}\label{tab4}
\end{center}
\end{table}

For the $P$ wave bottomonium case, see Table \ref{tab4}, similar to $P$ wave charnomium results, the relations $v^n_{\chi_{b0}(mP)}\approx v^n_{\chi_{b1}(mP)}\approx v^n_{\chi_{b2}(mP)}\approx v^n_{h_{b}(mP)}$ ($n,m=1,2,3,4$) are also exist. Except $v_{\chi_{bJ}(2P)}= v_{\chi_{bJ}(1P)}$ and $v_{h_{b}(2P)}\simeq v_{h_{b}(1P)}$, we have  $v^n_{\chi_{bJ}(4P)}>v^n_{\chi_{bJ}(3P)}>v^n_{\chi_{bJ}(2P)}>v^n_{\chi_{bJ}(1P)}$ and $v^n_{h_{b}(4P)}>v^n_{h_{b}(3P)}>v^n_{h_{b}(2P)}>v^n_{h_{b}(1P)}$ ($n=1,2,3,4$, $J=1,2,3$). Though we have the similar relations to charmonium system, that $v^n_{\chi_{bJ}(mP)}>v^n_{\Upsilon(mS)}$ ($n,m=1,2,3,4$, $J=1,2,3$) and $v^n_{h_{b}(mP)}>v^n_{\eta_{b}(mS)}$ ($n,m=1,2,3,4$), but different from the charmonium case, in a roughly estimation, we can choose the approximation $v^n_{\chi_{bJ}(mP)}\approx v^n_{h_{b}(mP)}\approx v^n_{\Upsilon(mS)}\approx v^n_{\eta_{b}(mS)}$ ($n,m=1,2,3,4$, $J=1,2,3$), for example, all the values of  $v^2_{\chi_{bJ}(1P)}$, $v^2_{h_{b}(1P)}$, $v^2_{\Upsilon(1S)}$ and $ v^2_{\eta_{b}(1S)}$ are around $0.071\sim0.076$, the existing of this relation is due to the very heavy bottom quark mass.

Reference \cite{buchmuller} considered the similar quantities, their values are $v^2_{b\bar b(1P)}=0.069$, $v^2_{b\bar b(2P)}=0.078$, and $v^2_{b\bar b(3P)}=0.090$, our values for a bottom inside $\chi_{b0}(nP)$ are $v^2_{\chi_{b0}(1P)}=0.076$, $v^2_{\chi_{b0}(2P)}=0.088$ and $v^2_{\chi_{b0}(3P)}=0.099$.
The values in Ref. \cite{buchmuller} are slightly smaller than ours, but two results are comparable. Ref. \cite{hwang} also predicted their results, which are $v^2_{b\bar b(1P)}=0.111$ and $v^4_{b\bar b(1P)}=0.0160$, larger than ours.

Because the bottom quark mass is very heavy, it moves slowly and has a small velocity in bottomonium, then its relativistic corrections are small, so the behavior of velocity expansion of the bottom quark if we make is much different from charm quark case. From Table \ref{tab3} and Table \ref{tab4}, we can see that the convergence in the velocity expansion is good, even for highly excited state. For example, we get $v_{\Upsilon(1S)}=0.24$, $v^2_{\Upsilon(1S)}=0.072$, $v^3_{\Upsilon(1S)}=0.025$, and $v^4_{\Upsilon(1S)}=0.010$, the convergence rate on the power of $v$ is much quick. The small values of $v^n$, and the good convergence in velocity expansion, indicate small relativistic corrections in bottomonium, including highly excited states.

We also note that in a quarkonium, as a expectation value, $\overline{v^2}\neq\bar{v}^2$, $\bar{v}\cdot \overline{v^2}\neq\bar{v}^3$ and $\overline{v^4}\neq\overline{v^2}^2$. For example in case of $J/\psi$, $\overline{v^2}=0.26$, but $\overline{v}^2=0.21$, $\overline{v^4}=0.14$ is much larger than $\overline{v^2}^2=0.067$. In case of $\Upsilon(1S)$, $\overline{v^2}=0.072\neq\overline{v}^2=0.058$, $\overline{v^4}=0.010\neq\overline{v^2}^2=0.0051$. We have the relation $\overline{v^n} >\overline{v^{n_1}}\cdot\overline{v^{n_2}}$, where $n_1+n_2=n$, and this relation is correct for all the charmonia and bottomonia. GK relation \cite{GK} predicted, $\langle v^{2n}\rangle=\langle v^2\rangle^n$, which is accurate up to corrections of order $v^2$. Our results show the deviation of GK relation from direct calculation, and the deviation is small when $n$ is small, but large when $n$ is large.

In summary, using the Bethe-Salpeter method, we calculate
the average values ${{q}^n}$ and $v^n$ ($n=1,2,3,4$) of $c$ and $b$ quarks in $S$ wave and $P$ wave quarkonia. We obtained, for example, $v_{J/\psi}=0.46$, $v^2_{J/\psi}=0.26$, $v^3_{J/\psi}=0.18$, $v^4_{J/\psi}=0.14$, and $v_{\Upsilon(1S)}=0.24$, $v^2_{\Upsilon(1S)}=0.072$, $v^3_{\Upsilon(1S)}=0.025$, $v^4_{\Upsilon(1S)}=0.010$. Our results also show the following relations, $v^n_{4S} > v^n_{3S}> v^n_{2S}>v^n_{1S}$, $v^n_{4P} > v^n_{3P}> v^n_{2P}>v^n_{1P}$, $v^n_{mP}>v^n_{mS}$ ($n,m=1,2,3,4$) and ${v^n} >{v^{n_1}}\cdot{v^{n_2}}$, where $n_1+n_2=n$.
We find highly excited states have larger relativistic corrections than those of the corresponding low excited and ground states, and the convergence of the velocity expansion is poor in charmonium system, especially bad for highly excited states, which indicate large relativistic corrections existing in charmonium system.

\section*{Acknowledgments}

This work was supported in part by the National Natural Science
Foundation of China (NSFC) under Grant No. 11575048, No. 11535002, No. 11625520.

\end{document}